\documentclass[12pt,hyper,a4paper]{JHEP3}
\usepackage{cite,graphicx,amsmath,amssymb,bbm}
\usepackage{epsfig}

\title{Flux Stabilization in 6 Dimensions: \\D-terms and Loop Corrections}
\author{A.~P.~Braun, A.~Hebecker and M.~Trapletti\\
Institut f\"ur Theoretische Physik, Universit\"at Heidelberg, 
Philosphenweg~16~und~19, D-69120 Heidelberg, Germany\\
\email{a.braun@thphys.uni-heidelberg.de,
a.hebecker@thphys.uni-heidelberg.de,
m.trapletti@thphys.uni-heidelberg.de}}


\newcommand{\be}{\begin{equation}}  
\newcommand{\ee}{\end{equation}}  
\newcommand{\bea}{\begin{eqnarray}}  
\newcommand{\eea}{\end{eqnarray}}

\renewcommand{\Im}{\operatorname{Im}}
\newcommand{\ol}[1]{\overline{#1}}
\newcommand{\ind}{\operatorname{ind}}

\abstract{
We analyse $D$-terms induced by gauge theory fluxes in the context of 
6-dimensional supergravity models. On the one hand, this is arguably
the simplest concrete setting in which the controversial idea of `$D$-term 
uplifts' can be investigated. On the other hand, it is a very plausible 
intermediate step on the way from a 10d string theory model to 4d 
phenomenology. Our specific results include the flux-induced one-loop 
correction to the scalar potential coming from charged hypermultiplets. 
Furthermore, we comment on the interplay of gauge theory 
fluxes and gaugino condensation in the present context, demonstrate explicitly 
how the $D$-term arises from the gauging of one of the compactification moduli, 
and briefly discuss further ingredients that may be required for the 
construction of a phenomenologically viable model. In particular, we show
how the 6d dilaton and volume moduli can be simultaneously stabilized, in the 
spirit of KKLT, by the combination of an R symmetry twist, a gaugino 
condensate, and a flux-induced $D$-term. }

\preprint{HD-THEP-06-30}
\begin{document}

\section{Introduction}
Fluxes are an essential ingredient in the compactification of 
string-theoretic and other higher-dimensional models (see~\cite{ 
Douglas:2006es} for a recent review). In the present paper, we analyse the 
effects that $D$-terms induced by gauge theory fluxes can have in the context 
of moduli stabilization of 6d supergravity compactified to 4 dimensions. 
Although, at a technical level, the paper is entirely field-theoretic and 
relies only on the familiar 6d supergravity lagrangian of~\cite{ 
Nishino:1986dc}, our motivation is largely string-theoretic, as we now 
explain: 

One of the perceived problems of the KKLT construction~\cite{Kachru:2003aw} 
of metastable de-Sitter vacua of type IIB supergravity is the presence of 
$\ol{\rm D3}$-branes (`anti-D3-branes'), which break supersymmetry 
explicitly. Before supersymmetry breaking, the model is characterized by the 
superpotential 
\begin{equation}
W=W_{0}+Ae^{-aT}\label{Wkklt}\,,
\end{equation} 
where $T$ is a chiral superfield with no-scale K\"ahler potential. The AdS 
vacuum of this model is then `uplifted' by adding $\ol{\rm D3}$-branes in a 
strongly warped region. Since no ${\cal N}=1$ supergravity description of this 
construction has so far been derived from first principles, their effect is 
usually incorporated by adding an uplifting term~\cite{Kachru:2003aw,
Kachru:2003sx}
\be
V_{\ol{\rm D3}}\sim\frac{1}{(T+\ol{T})^{2}}
\ee 
directly to the scalar potential (i.e. without specifying the modified 
supergravity model).\footnote{
It 
has, however, been argued that a phenomenologically motivated description in 
terms of non-linearly realized supersymmetry is sufficient for most practical 
purposes~\cite{Choi:2005ge}. This is consistent with the observation that, 
when modelling the $\ol{\rm D3}$ brane uplift by $F$-term breaking, the 
phenomenology turns out to be independent of the detailed dynamics of this 
SUSY breaking sector (unless extra fields 
violate the underlying sequestering assumption)~\cite{Brummer:2006dg}.}

Following Burgess, Kallosh and Quevedo~\cite{Burgess:2003ic}, one can attempt 
to avoid these difficulties by introducing, instead of $\ol{\rm D3}$-branes, 
supersymmetry-breaking two-form flux on the worldvolume of D7-branes. This 
has a well-known ${\cal N}=1$ supergravity description in terms of a 
supersymmetry-breaking $D$-term potential  (see 
e.g.~\cite{Kachru:1999vj,Jockers:2005zy,Villadoro:2006ia,Haack:2006cy})
\begin{equation}
V_{D}\sim\frac{D^2}{T+\overline{T}}\sim\frac{1}{(T+\overline{T})^{3}}
\hspace{.1ex}\,.
\end{equation}

As emphasized by a number of authors~\cite{Dudas:2005vv,Choi:2005ge,
Villadoro:2005yq,deAlwis:2006sz}, there are, however, two fundamental problems
with this proposal: one related to the intimate connection between $F$- and 
$D$-terms, the other to the gauge invariance of the superpotential. The second 
problem becomes apparent if one recalls that $D$-terms originate from the 
gauging of an isometry of the scalar manifold of the supergravity model~\cite{ 
WessBagger}. In the present case, the relevant symmetry is the shift symmetry 
acting on the imaginary part of $T$ (see~\cite{Jockers:2005zy} for a detailed 
discussion). However, the superpotential of Eq.~(\ref{Wkklt}) is not invariant 
under a shift in Im$(T)$, rendering the whole construction inconsistent. To
be more precise, while a transformation of $W$ by a complex phase could be
tolerated (being equivalent to a K\"ahler-Weyl transformation), the presence 
of the constant $W_0$ induces a more complicated behaviour and thus an actual 
inconsistency. Thus, there is a clash between three ingredients: the 
3-form-flux-induced constant $W_{0}$, the gaugino condensate inducing 
the exp$(-aT)$ contribution \cite{gauginos}, and the gauging of the shift 
symmetry in $\Im(T)$. 
Any two of these three ingredients may be able to coexist in a consistent 
model.

The above clash can obviously be avoided if light fields other than $T$ are 
present and the coefficient $A$ of the exponential term in Eq.~(\ref{Wkklt})
depends on them in such a way as to render $W$ gauge invariant. Indeed, 
this is well-known to occur in 4d supersymmetric gauge theory~\cite{ 
Affleck:1983mk} (see~\cite{Dudas:2005vv,Achucarro:2006zf} for a discussion
in the present context).
It has very recently been demonstrated~\cite{Haack:2006cy} that, 
as expected, consistent type IIB compactifications avoid any potential 
inconsistency between 2-form flux and gaugino condensation by precisely this 
mechanism.\footnote{For a discussion of the related clash between the 
gauging of isometries and superpotential corrections by instantons 
see~\cite{Kashani-Poor:2005si}.}

However, this resolution of the gauge-invariance problem has serious 
implications for the whole stabilization/uplifting proposal. To see this,
let $A=A(\Phi_1,\ldots,\Phi_n)$ and formulate the gauge-invariance 
requirement for $W$ as
\be
(X^{\Phi_i}\partial_{\Phi_i}+X^T\partial_T)W=0\,,
\ee
where $X$ is the Killing vector of the isometry to be gauged. We can now 
re-parameterize the scalar manifold as follows: Choose some complex 
$n$-dimensional submanifold which is nowhere parallel to $X$, parameterize it 
arbitrarily by $n$ variables, and finally parameterize the motion of 
this manifold along $X$ by one last variable $z$. Clearly, in this 
parameterization $X^z$ is the only non-zero component of the Killing vector 
and $W$ is independent of the $z$ superfield. Thus, in the 
original AdS vacuum, $D_zW=K_zW=0$ and hence $K_z=\partial K/\partial z=0$. 
This implies that the $D$-term arising after the gauging of the $X$-isometry, 
$D=iK_zX^z$, automatically vanishes at the point of the AdS vacuum.\footnote{
This
last statement is a simplified rendition of the above-mentioned 
$F$-term/$D$-term problem. It can, in principle, be avoided by allowing for 
a Fayet-Iliopoulos term (an additive constant contribution to $D$ which is
not proportional to $K_z$). However, such effects do not arise the present 
context of 2-form-flux-induced $D$-terms. 
Furthermore, the above does clearly not represent an objection to the
$D$-term uplift \cite{Parameswaran:2006jh} of non-SUSY AdS vacua
\cite{Berg:2005yu} arising from the interplay of $\alpha^\prime$ corrections
\cite{Becker:2002nn} and K\"ahler corrections.}

Does this mean that the AdS vacuum of KKLT can not be uplifted by a small
correction related to the gauging of an isometry? We would like to argue the
opposite as follows. The $D$-term uplift has to be understood as a twofold 
modification of the model in which the AdS vacuum occurs: One ingredient is 
the inclusion of extra light fields ($A \rightarrow A(\Phi)$ in the simplest 
case), the other is the gauging of an isometry. Without loss of generality, 
we can assume
\be 
W=W_0+Ae^{-a(\Phi+T)}\,,
\ee
with $A$ an appropriately redefined constant. Furthermore, we can make the 
ansatz
\be
K=-3\ln(T+\ol{T})+g(\Phi+\ol{\Phi})
\ee
for the K\"ahler potential, which allows us to gauge the isometry $T\to 
T+i\delta$, $\Phi\to\Phi-i\delta$. Allowing ourselves to choose an arbitrary 
functional form for $g$ it is clearly possible to arrange for the scalar 
potential
\be
V=e^K\left[K_{\Phi\ol{\Phi}}^{-1}|D_\Phi W|^2+K_{T\ol{T}}^{-1}|D_T W|^2-
3|W|^2\right]+\frac{1}{2}({\rm Re}h)^{-1}D^2
\ee
to have a minimum near the original AdS vacuum with small $F$ terms and a 
sizeable $D$-term $D$. (Here $h$ is the gauge-kinetic function.) This 
might then be viewed as a $D$-term uplift of the original SUSY AdS vacuum. 

Moreover, the following can be viewed as a limiting case of the above 
proposal: Leave the $T$ sector of the model, responsible for the SUSY AdS
vacuum, completely unchanged (avoiding in particular any attempt to gauge the 
shift in Im$T$). Instead, add an extra superfield $\Phi$ and gauge it in such 
a way that, in the vacuum, the $D$-term dominates over the $F$ -term. One 
might consider such an approach as a $D$-term analogue\footnote{
Phenomenological constraints on non-sequestered $D$-term uplifts were
discussed in~\cite{Choi:2006bh}.} 
of uplifts by non-linearly realized SUSY~\cite{Choi:2005ge}, by $F$-terms 
in the strongly warped region~\cite{Brummer:2006dg}, or by dynamical SUSY
breaking \cite{Dudas:2006gr} 
(see also~\cite{Luty:2002hj,Lebedev:2006qq}). 
It is not known whether this or the previously 
outlined variant of a $D$-term uplift have a string-theoretic realization, 
but there appear to be no fundamental inconsistencies. 

In our following investigation of 6d supergravity with 2-form-flux 
\cite{Salam:1984cj,Aghababaie:2002be,gibbons} (see also
\cite{Lee:2005az} for recent related work),
we will not be able to realize one of these conceivable scenarios to full 
satisfaction. However, we will develop a number of ingredients that may 
be useful in this pursuit in the future. 

We begin in Sect.~2 by analysing in detail a simple $T^2/Z_{2}$ model
(easily generalizable to $T^2/Z_n$) in which two modulus superfields $S$ and 
$T$ encode (different combinations of) the dilaton and the compactification 
volume. We calculate the scalar potential arising in the presence of 
2-form-flux in two ways -- by integrating the $F_{56}^2$ term over the compact 
space and by finding the $D$-term that arises from the gauge transformation of 
$T$. Since the superfield $S$, which governs all gauge-kinetic 
functions, does not transform, no gauge invariance problem arises in the 
presence of gaugino condensation.\footnote{A related situation occurring in
the presence of both flux and gaugino condensation on the {\it same}
D7-brane-stack is discussed in the Appendix of~\cite{Haack:2006cy}.} 

We continue in Sect.~3 by calculating the one-loop correction to the scalar 
potential that arises if hypermultiplets charged under the fluxed U(1) are
present. Its parametrical behaviour is that of a usual Casimir energy, i.e. 
$\sim 1/R^4$ in the Brans-Dicke frame (the frame where the coefficient of the 
4d Einstein-Hilbert term is proportional to the torus volume $R^2$). Due to the
quantized coefficient of this loop correction, it is potentially more important
than Casimir energies induced by other (weak) SUSY breaking effects. 

One such SUSY breaking effect, which we discuss in Sect.~4, is 6d 
Scherk-Schwarz breaking. In close analogy to the more familiar 5d case, it 
is implemented using an SU(2)$_R$-symmetry twist and can be viewed, from the 
4d perspective, as the introduction of a constant superpotential $W_0$. We 
also comment on the (im-)possibility of this type of SUSY breaking on $Z_n$ 
orbifolds with various $n$ and on other mechanisms for the generation of a 
non-zero superpotential.

In Sect.~5 we discuss options for moduli stabilization using the various 
ingredients analysed above. Working on a $T^2/Z_2$ orbifold and ignoring, 
for simplicity, the shape modulus of the torus, one still has to deal with 
the stabilization of the superfields $S$ and $T$ simultaneously. At fixed $T$, 
the modulus $S$ is stabilized \`a la KKLT by the interplay of $W_0$ and 
gaugino condensate. The depth of the resulting SUSY AdS vacuum depends on 
$T$, driving Re$T$ to small values. This is balanced by the $T$
dependence of the flux-induced $D$-term, leading to a stable non-SUSY AdS 
vacuum. Thus, while the 2-form flux does not provide the desired uplift,
it plays an essential role in the simultaneous stabilization of two moduli. 
Unfortunately, the loop correction has the same $T$ dependence as the flux 
term (being suppressed by an extra power of Re$S$) so that an uplift using 
the former is impossible (at least within our step-by-step approximate 
analysis). However, we consider the possibility of a simultaneous stabilization
of two moduli by the interplay of $W_{0}$, gaugino condensate and $D$-term an
interesting and positive result. The required uplift can, in the present 
context, be provided by $F$-terms of the $\mathcal{N}=1$ sectors localized
at the orbifold fixed points.

Our conclusions are given in Sect.~6 and some technical details of the loop
calculation are relegated to the Appendix.

\section{A six-dimensional model}
We work with the following bosonic action for supergravity coupled to gauge 
theory in six dimensions~\cite{Nishino:1986dc,Salam:1984cj}
\footnote{We use the conventions of Appendix~B of~\cite{polchinski}. 
Note that our action 
contains a tensor multiplet besides the supergravity and the vector multiplet. 
If one wants to work with a Lorentz invariant action this enlargement of the 
minimal setup is unavoidable\cite{Marcus:1982yu}.
}:
\begin{equation}
\sqrt{-g_{6}}^{\hspace{.5ex}-1}\mathcal{L}=
-\frac{1}{2}\mathcal{R}_{6}-\frac{1}{2}\partial_{M}
\phi\partial^{M}\phi-\frac{1}{24}e^{2\phi}H_{MNP}H^{MNP}
-\frac{1}{4}e^{\phi}F_{MN}F^{MN} \label{L}.
\end{equation}
The field strength $H$ is defined as
\begin{eqnarray}
\label{defH}
H_{MNP}=
\partial_{M}B_{NP}+F_{MN}A_{P}+\mbox{cyclic permutations}=
(dB+F\wedge A)_{MNP}
\end{eqnarray} 
and the above action is invariant under the gauge transformations
\begin{equation}
\delta A=d\Lambda\,,\,\,\,\,\,\,
\delta B=-\Lambda F+dC\,.\label{Btrafo}
\end{equation}
The extra symmetry related to the Kalb-Ramond $B$-field and implemented by 
the $\mbox{1-form}$ $C$ will be crucial in the presence of fluxes for $F$. 
The metric of the six-dimensional spacetime $R^4\times T^2$ is taken to be
\begin{equation}
(g_{6})_{MN}= \left(\begin{array}{cc} r^{-2}(g_{4})_{\mu\nu } & 0 \\ 0 & 
r^{2}(g_{2})_{mn} \end{array}\right)\,,\label{metric}
\end{equation}
with $\mu,\nu = 0 .. 3$ and $m,n= 5 .. 6$. The $r^{2}$ in front of 
$(g_{2})_{mn}$ controls the size of the extra dimensions in a convenient 
fashion, whereas the $r^{-2}$ in front of $(g_{4})_{\mu\nu }$ acts as an 
automatic Weyl rescaling to ensure that the Einstein-Hilbert term in 4D is 
canonical. The metric of the internal space is
\begin{equation}
\label{intmet}
(g_{2})_{mn}=
\displaystyle\frac{1}{\tau_{2}}\left(
	\begin{array}{cc} 
	1 & \tau_{1} \\
	\tau_{1} & \tau_{1}^{2}+\tau_{2}^{2} 
	\end{array}\right)\,,
\end{equation}
with the modulus $\tau\equiv\tau_{2}+i\tau_{1}$ controlling the shape of the
torus. The domain of $x_{5}$ and $x_{6}$ is taken to be a square of unit 
length, so that $\int\sqrt{g_{2}}d\hspace{-.2ex}x^{5}d\hspace{-.2ex}x^{6}=1$. 

We introduce a constant background for the field strength $\langle F_{mn} 
\rangle=f\epsilon_{mn}$, with $f$ a quantized number, as typically required 
in a string model. We split the gauge potential $A$ into a fluctuation term 
$\mathcal A$ and a background term $\langle A\rangle$, such that $\langle 
F\rangle=d\langle A\rangle$. The background $\langle A\rangle$ cannot be 
globally defined in the internal space. On the overlap of different patches, 
background gauge transformations with a parameter $\Lambda_0$ are required: 
\begin{equation}
\delta_{\Lambda_0} \langle A\rangle=d\Lambda_0,\,\,\,\,\,\,
\delta_{\Lambda_0} \mathcal A=0\,.
\end{equation}
Given the general gauge transformation formulae
\begin{equation}
\delta_{\Lambda_0} A=d\Lambda_0,\,\,\,\,\,\,
\delta_{\Lambda_0} B=-\Lambda_0 F+dC\,,
\end{equation}
it follows that also $B$ is not globally defined, since it is not possible 
to absorb $-\Lambda_0 F$ in  $dC$. This is clear since the variation of 
$dB$, which is independent of $C$, is in general non-trivial:
\begin{equation}
\delta_{\Lambda_0} dB=-d\Lambda_0 \wedge F\,.
\end{equation}
The last expression can be rewritten according to
\be
\delta_{\Lambda_0} dB=-d\Lambda_0\wedge (\langle F\rangle + d{\cal A})=
-d\Lambda_0\wedge d{\cal A}=d(d\Lambda_0\wedge{\cal A})
=\delta_{\Lambda_0} d\left(\langle A\rangle\wedge \mathcal A\right)\,,
\ee
which shows that, for a new field $\mathcal B=B-\langle A\rangle \wedge 
\mathcal A$, the quantity $d{\cal B}$ is globally defined. Moreover, the new 
2-form ${\cal B}$ will itself be globally defined provided that 
\begin{equation}
\delta_{\Lambda_0} \mathcal B=-\Lambda_0 F-d\Lambda_0\wedge\mathcal A+dC=0\,.
\end{equation}
The required 1-form $C=C(\Lambda_0,{\cal A},\langle A\rangle)$
(satisfying $dC=\Lambda_0 F+d\Lambda_0\wedge\mathcal A$) can indeed be
explicitly given in the case of constant background flux~\cite{Kaloper:1999yr}.

In conclusion, all the degrees of freedom of $B$ are now described by a new 
field $\mathcal B=B-\langle A\rangle \wedge\mathcal A$, that is globally
defined in the internal dimensions, and thus has a standard Kaluza-Klein
expansion. The gauge transformations of ${\cal B}$ follow from its definition 
together with Eq.~(\ref{Btrafo}) and the explicit form of $C$. They simplify 
if we focus on $d\mathcal B$ since $C$ drops out:
\begin{equation}
\delta d\mathcal B=-2 d\Lambda\wedge \langle F\rangle-d\Lambda\wedge 
d\mathcal A\,.
\end{equation}
For 4d gauge transformations $\Lambda=\Lambda(x^\mu)$, this can be written as
\begin{equation}
\delta\left( \partial_\mu {\cal B}_{56}+\partial_5 {\cal B}_{6\mu}+\partial_6 
{\cal B}_{\mu 5}\right)=-2\partial_\mu \Lambda \langle F_{56}\rangle-
\partial_\mu \Lambda (\partial_5 \mathcal A_6-\partial_6 \mathcal A_5)\,.
\end{equation}
Restricting ourselves to the zero-mode level, any dependence of the internal 
coordinates drops out and we find
\begin{equation}
\label{gaugvar}
\delta  \mathcal B_{56}=-2 \Lambda \langle F_{56}\rangle
\end{equation}
for the ${\cal B}_{56}$ zero mode. Note the factor-of-two difference from 
the naive expectations that one might have for $B_{56}$ on the basis of
Eq.~(\ref{Btrafo}).\footnote{We thank Giovanni Villadoro for discussions
about this issue.}
They are not justified since $B$ is not globally defined on the internal space 
and possesses no standard Kaluza-Klein expansion. 

We will be interested in the 4d theory arising from the compactification on a
supersymmetric $T^2/Z_2$ orbifold (see~Sect.~4 for details). Hence we 
disregard all 4d vector multiplets which are eliminated by the orbifold 
projection, as well as the Wilson line degrees of freedom associated with 
the 5d U(1) gauge theory. What remains are the 4d supergravity and the vector 
multiplet with gauge field $A_\mu$ together with three chiral multiplets, the 
moduli of the compactification. The latter contain the degrees of freedom 
$r$, $\phi$, $\tau_{1}$, $\tau_{2}$ and two scalars related to the 
2-form $\mathcal B$. The lowest components of the three modulus superfields 
are~\cite{Aghababaie:2002be,Falkowski:2005zv}
\begin{equation}
S\equiv\tfrac{1}{2}(s+ic),\quad T\equiv\tfrac{1}{2}(t+ib),
\quad \tau\equiv\tfrac{1}{2}(\tau_{2}+i\tau_{1})\label{defTStau}.
\end{equation}
where we have used the definitions
\be
t\equiv e^{-\phi}r^{2}\,,\qquad s\equiv e^{\phi}r^{2}
\ee
and
\be
b\epsilon_{mn}\equiv \mathcal B_{mn}\,,\qquad
\epsilon_{\mu\nu\rho\sigma} \partial^{\sigma}c\equiv 
r^{4} e^{2\phi}(d\mathcal B)_{\mu\nu\rho}\,.
\ee
The K\"ahler potential, which can be inferred from the kinetic terms for the 
scalars after dimensional reduction and Weyl rescaling, reads 
\begin{equation} 
K=-\log(T+\overline{T})-\log(S+\overline{S})-\log(\tau+\overline{\tau})\,.
\end{equation} 
Similarly, the gauge-kinetic function is found to be $h(S)=2S$ (using the 
standard conventions of~\cite{WessBagger}).

Given Eq.~(\ref{gaugvar}), the 4d gauge transformations read
\be
\delta b=-2f\Lambda,\hspace{.5cm}\delta \mathcal A_{\mu}=\partial_{\mu}\Lambda,
\ee
which implies that the only nonvanishing component of the Killing vector
is $X^{T}=-if$. The resulting $D$-term $D=iK_TX^T=-f/t$ leads to the $D$-term
potential
\be
V_{D}=\frac{f^{2}}{2st^{2}}\,.
\ee
The same potential also follows directly from the 6d gauge-kinetic
term, evaluated in the flux background: 
\be
\int d^{6}\hspace{-.2ex}x\sqrt{g_{6}}\,
\frac{e^{\phi}}{4}\langle F_{MN}F^{MN}\rangle=
\int d^{4}\hspace{-.2ex}x\sqrt{g_{4}}\,e^{\phi}\frac{f^{2}}{4r^{6}}
\epsilon_{mn}\epsilon^{mn}=\int d^{4}\hspace{-.2ex}x\sqrt{g_{4}}
\frac{f^{2}}{2st^{2}}\,.
\ee
This represents a nontrivial check of the fact that the flux is described by 
the gauging of an isometry from the 4d perspective. (See~\cite{Villa} 
for a similar computation in heterotic string theory.)
Note in particular that, as advertised in the introduction, the gauge 
transformation acts only on $T$, while the gauge kinetic function depends
only on $S$. Hence, no clash between gaugino condensate and $D$-term potential
arises. A related situation occurring in the presence of both flux and gaugino 
condensation on the {\it same} D7-brane-stack has recently been discussed 
in the Appendix of~\cite{Haack:2006cy}.

\section{Loop corrections}
As an example of a loop correction in the presence of flux, the one-loop
Casimir energy of a charged 6d hypermultiplet is computed in this section. 
This is expected to be the dominant contribution because the constituents of 
the charged hypermultiplet feel the flux directly. We first derive the Casimir 
energy for $T^{2}$ and then redo the computation with the degrees of freedom 
that remain in the spectrum for $T^{2}/Z_{2}$. \par

The constraints on the gauge and matter content of a
consistent anomaly free 6d theory \cite{Salam:1985mi} allow the presence of the
charged hypermultiplets that we are introducing.
Unfortunately, these constraints typically impose also the presence of extra
gauge sectors, with extra matter multiplets, whose analysis goes beyond
the scope of the present work.
In this sense, our model has to be considered as a sector of a complete theory,
under the assumption that such a completion does not affect the moduli 
stabilization studied here.

For the Casimir energy calculation one first has to derive the mass 
spectra of the charged 6d scalars and Weyl fermions. A 6d hypermultiplet 
consists of two complex scalars and one 6d Weyl fermion which enter the action
in a quite complicated way~\cite{Nishino:1986dc}. We will linearize the 
$\sigma$-model and work with canonical kinetic terms, neglecting the 
self-interactions of the scalars. This is expected to be a good approximation
as long as the mass scale of gauge interactions in 6d is much lower than the 
6d~Plank~scale, $1/g_{\rm YM,6}\ll M_{\rm Pl,6}$.
Note that the kinetic terms do not 
contain the 6d dilaton $\phi$~\cite{Nishino:1986dc}. In the derivation of the
mass spectra we follow \cite{Bachas:1995ik}.
\par
As in the case without flux, the masses of the scalars are given by the 
eigenvalues of the Laplacian on the compact space. For one minimally 
coupled complex scalar field with covariant derivative $\mathcal{D}$, the 
Laplacian reads
\begin{equation}
\frac{1}{r^{4}}\left(\mathcal{D}_{5}^{2}+\mathcal{D}_{6}^{2}\right),
\end{equation}
where we have used the decomposition of Eqs.~(\ref{metric})
and~(\ref{intmet}), assuming $\tau_1=0$ and $\tau_2=1$. 
In the case of a nonzero constant flux the covariant derivatives no longer
commute,
\begin{equation}
\left[\mathcal{D}_{5},\mathcal{D}_{6}\right]=iF_{56}=if.
\end{equation}
Algebraically, this is equivalent to a one-dimensional harmonic oscillator
with unit mass and unit frequency. For positive $f$ the correspondence is 
\bea
\mbox{Hamiltonian}&\leftrightarrow& \tfrac{1}{2}\left(\mathcal{D}_{5}^{2}+
\mathcal{D}_{6}^{2}\right)\nonumber\\
\mbox{position}&\leftrightarrow& \mathcal{D}_{5}\nonumber\\
\mbox{canonical momentum}&\leftrightarrow& \mathcal{D}_{6}\nonumber\\
\hslash&\leftrightarrow& f.
\eea
For negative $f$, the position and momentum operators have to be interchanged
but the mass spectrum is not affected. It reads
\begin{equation}
m_{n}^{2}=\frac{2\vert f\vert}{r^{4}}\left(n+\tfrac{1}{2}\right),\label{mbos}
\end{equation}
where $n$ is a non-negative integer.
Note that the $n$-dependence of this mass spectrum is quite different from the 
usual Kaluza-Klein tower ($m^{2}\sim n_1^{2}+n_2^{2}$) resulting from compact 
dimensions without flux. \par

Some care has to be taken in deriving the fermionic Kaluza-Klein towers, as is 
explained in Chapter 14 of~\cite{Green:1987mn}. Since the Dirac operator 
couples righthanded fermions to lefthanded fermions, only its square can have 
eigenfunctions. The masses of the fermions are determined by
\begin{equation}
m_{n}^{2}r^{4}\psi_{n}=\left(\Gamma^{5}\mathcal{D}_{5}+\Gamma^{6}\mathcal{D}_{6}\right)^{2}\psi_{n},
\end{equation}
where the $\psi_{n}$ are 6d spinors. 
Observing that
\begin{equation}
\left(\Gamma^{5}\mathcal{D}_{5}+\Gamma^{6}\mathcal{D}_{6}\right)^{2}=
\mathcal{D}_{5}^{2}
+\mathcal{D}_{6}^{2}+i\Gamma^{5}\Gamma^{6}f,
\end{equation}
it is clear that the problem differs from the bosonic case only by a shift
if the spinors are eigenvectors of $\Gamma^{5}\Gamma^{6}$.
To quantify the effect of the shift, recall that
\begin{equation}
\Gamma^{7}=i\gamma^{5}\Gamma^{5}\Gamma^{6},
\end{equation}
and that the 6d chirality is fixed. Decomposing the 6d spinor into a
direct sum of two 4d Weyl spinors, we now see that the shift is determined by 
the chirality of each 4d spinor. The fermionic eigenfunctions are the same as 
the bosonic ones, the only difference is that they carry an extra
chirality index which induces a shift of their masses. 
The mass spectrum of 4d 
Weyl fermions reads 
\begin{equation}
(m_{n}^{2})_{\pm}=\frac{2\vert f\vert}{r^{4}}
\left(n+\tfrac{1}{2}\pm\tfrac{1}{2}\right).\label{mferm}
\end{equation} 
\par
Another point which has to be addressed is the degeneracy of the spectra. The 
quickest derivation uses the two-dimensional index theorem, which in our case 
counts the number of massless fermions. We find that
\begin{equation}
\ind(\Gamma^{5}\mathcal{D}_{5}+\Gamma^{6}\mathcal{D}_{6})=\frac{1}{2\pi}
\int_{T^{2}}F=\frac{f}{2\pi}=N\label{index}.
\end{equation}
\par
Thus the monopole number equals the degeneracy of the state with vanishing 
mass. It is clear that the ground state of the fermions of opposite chirality 
has the same degeneracy, because we are considering the same Laplace operator
to which merely a constant is added, and thus we find precisely the same
eigenfunctions.
By the same argument we conclude that the bosonic ground state is $N$-fold
degenerate.\footnote{This can also be checked by explicitely computing the
zero eigenfunctions. They are given in the Appendix.}
From the oscillator~algebra it then follows that all excited states
have the same degeneracy as the ground states. Thus every fermionic and every 
bosonic level is populated by $N$ states. An extra factor of two
arises in the bosonic sector because of the two complex scalars in the 
hypermultiplet.
\par
With this particle spectrum we directly compute the one-loop effective
potential from a four-dimensional perspective. In dimensional regularisation 
and after Wick rotation to Euclidean space it reads
\cite{Peskin:1995ev}:
\be
\sum_{\delta=0,\hspace{.2ex}\pm 1/2}(-1)^{2\delta}(2-2\vert\delta\vert)
\vert N\vert\sum_{n=0}^{\infty}\int \frac{d^{D}\hspace{-.2ex}k}{(2\pi)^{d}}\ln(k^{2}+
m_{n}^{2}(\delta)),
\ee 
where
\be
m_{n}^{2}(\delta)=\frac{2\vert f\vert}{r^{4}}(n+\tfrac{1}{2}+\delta)
\ee
are the bosonic ($\delta=0$) and fermionic ($\delta=\pm 1/2$) mass spectra. 
This expression is computed in the Appendix, giving the result:
\begin{eqnarray}
V_{\mbox{\scriptsize{Casimir}}}&=&
\frac{7}{4}\frac{\vert N\vert^{3}}{(st)^{2}}\zeta_{R}'(-2)
\cong-0.053\frac{1}{(2\pi)^{3}}\frac{\vert f\vert^{3}}{(st)^{2}}\label{VC-no-z2}   
\end{eqnarray}
Here we have used the quantization condition for the flux, Eq.~(\ref{index}).
\par

The computation is analogous, albeit technically more involved, 
in the $T^{2}/Z_{2}$ case. Details are presented in the 
Appendix. The result is: 
\bea
V_{\mbox{\scriptsize{Casimir}}}^{\pm}&=&\frac{7}{4}\frac{1}{(2\pi)^2}
\left(\frac{f}{st}
\right)^{2}\zeta_{R}'(-2)J_{N}^{\pm}\nonumber\\
&=&-0.053\frac{1}{(2\pi)^{2}}\left(\frac{f}{st}\right)^{2}J_{N}^{\pm},
\eea
where we have defined
\be
J_{N}^{\pm}\equiv \vert N\vert\pm 4
\ee
and the flux quantization condition is
\begin{equation}
\frac{1}{2\pi}
\int_{T^{2}/Z_2}F=\frac{f}{4\pi}=N,
\end{equation}
where the first equality follows from the definition of $T^2/Z_2$,
the second one from the fact that the flux quantization condition on a
sphere is equal to that on a torus.

The two signs in $V^{\pm}$ stem from the different
internal parity that may be assigned to the fermions on the
massless level. 

This correction should be understandable as a correction to the 
K\"ahler potential.
We found a non-zero Casimir energy because SUSY is broken, which in turn is a
result of the flux. The flux was shown to generate a $D$-term potential in 
Sect.~2. We can trace the correction to the $D$-term potential back to a 
correction to the K\"ahler potential if we assume that the gauge symmetries
of our model remain unchanged. Neglecting higher orders in $1/r$ we find 
\be
\frac{f^{2}}{st}(\Delta K)_{T}
=-\frac{1}{(2\pi)^{2}}\frac{7}{4}\zeta_{R}'(-2)
\left(\frac{f}{st}\right)^{2}J_{N}^{\pm},
\ee
so that we can conclude
\be
\Delta K=-\frac{1}{(2\pi)^{2}}\frac{7}{4}\zeta_{R}'(-2)
\left(\frac{1}{S+\ol{S}}\log(T+\ol{T})\right)J_{N}^{\pm}.
\ee

\section{Scherk-Schwarz twists as a source for $W_0$}
The presence of closed string fluxes in a type IIB model induces a
superpotential $W_{flux}(z)$, that depends on the complex structure moduli 
$z$. The latter are thus stabilized at certain 
values $z_{min}$ and, from the point of view of the low-energy effective 
theory, the superpotential at the minimum is a constant $W_0=W_{flux}( 
z_{min})$. If $W_0\neq 0$, a SUSY-breaking no-scale model results. In the 
KKLT construction, a SUSY AdS vacuum is present due to the interplay between 
$W_0$ and gaugino condensation. We would like to reproduce this basic 
structure in our 6d approach. We could in principle introduce a constant 
$W_0$ in our model by appealing to the presence of closed string fluxes, 
since the model can be seen as an intermediate step in the compactification 
of 10d string theory. In praxis this is not convenient for the following 
reason. If, on the one hand, closed string fluxes are present in the 6d bulk 
we consider, we have to start from a more complicated lagrangian. If, on the 
other hand, the relevant fluxes are present only in the ``hidden'' four extra 
dimensions, we loose much of the explicitness of our construction, which is
based on a well-known consistent 6d supergravity model. It is therefore 
convenient to introduce $W_0$ as the manifestation of Scherk-Schwarz (SS) 
twists in the two compact extra dimensions as we now discuss in more 
detail~\cite{SS} (see~\cite{Lee:2005tk} specifically for the 6d case). 

The 6d supergravity theory studied in Sect.~2 possesses an SU(2)$_{\rm R}$
R-symmetry. This can be checked by direct inspection, or by considering it
as the result of the compactification of 10d string theory~\cite{
Imamura:2001es}. We follow the second approach. A 10d Majorana-Weyl 
spinor (a real {\bf 16} of SO(1,9)) transforms as ${\bf 4\oplus 4'}$ under 
the SO(1,5) subgroup. The action of the R symmetry group SU(2)$_{\rm R}\times
$SU(2)$_{\rm R'}$, which comes from SO(1,9)$\,\supset\,$SO(1,5)$\times$SO(4)$
\,=\,$SO(1,5)$\times$SU(2)$_{\rm R}\times$SU(2)$_{\rm R'}$, is such that the 
{\bf 4} and ${\bf 4'}$ transform only under SU(2)$_{\rm R}$ and under 
SU(2)$_{\rm R'}$ respectively.\footnote{
Notice 
that the R-symmetry group does not mix spinors with different 6d
chirality. Indeed, SO(4) can change neither the internal chirality,
by definition, nor the 10d chirality, since it is part of SO(1,9), and
the product of 6d and internal chirality gives precisely the 10d chirality.

A slightly different perspective on the situation can be given as follows:
A 10d Weyl spinor (a complex {\bf 16}) transforms under SO(1,5)$\times 
$SU(2)$_{\rm R}\times$SU(2)$_{\rm R'}$ as ({\bf 4},{\bf 2},{\bf 1})$\,\oplus$
(${\bf 4'}$,{\bf 1},{\bf 2}). The 10d reality constraint is imposed on each 
of these two terms independently, without mixing them. This leads to two 
complex-4-dimensional representations of both SO(4) and SU(2) which, 
however, can not anymore be viewed as a $({\bf 4},{\bf 2})$, i.e., as a 
tensor product of two complex representations. 
}
Consider now the compactification of a 10d ${\cal N}=1$ model on some 
orbifold limit of $K3$, such as $T^4/Z_n$. The SUSY generator is a real 
{\bf 16}. Taking the orbifold group to be generated by one of the elements 
of SU(2)$_{\rm R'}$, the supersymmetry associated with the ${\bf 4^\prime}$ is 
broken, while that associated with the {\bf 4} is preserved. We thus end up 
with the R-symmetry group SU(2)$_{\rm R}$ since, as explained, the {\bf 4} is 
also a doublet of SU(2)$_{\rm R}$. 

In the presence of the SU(2)$_{\rm R}$ symmetry, we can compactify the 6d 
theory on $T^2$ imposing non-trivial field-identifications. Given a generic 
SU(2)$_{\rm R}$ doublet $\Phi(x^\mu,x^5,x^6)$ (e.g. the gaugino)
we require
\bea
\Phi(x^\mu,x^5,x^6)=T_5 \Phi(x^\mu,x^5+1,x^6),\,\,\,
\Phi(x^\mu,x^5,x^6)=T_6 \Phi(x^\mu,x^5,x^6+1),
\eea
where the matrices $T_i$ embed the translations $t_i$ along the torus
coordinate $x^i$ in the R-symmetry group. Since $t_5 t_6=t_6 t_5$, we also 
require $T_5 T_6=T_6 T_5$. In case one (or both) of the matrices are 
non-trivial, we obtain a SS dimensional reduction. If one of the two matrices 
is trivial, e.g. $T_6$, the consistency requirement is automatically 
satisfied and we can shrink the $x^6$ direction, obtaining an effective 5d 
model. From this perspective, the SS twist due to $T_5$ can be seen as a 
standard SS twist in a 5d model compactified on $S^1$. 

For an orbifold compactification of the 6d theory, the rotation
operator $r\!\!\in\,$SO(2) is also embedded in the R-symmetry group via a 
matrix $R$. A non-trivial embedding is crucial for SUSY not to be broken in a 
hard way: in case $R={\mathbbm 1}$ the net action of the orbifold on any 4d
spinor would indeed result in a non-trivial phase, projecting it out of the
spectrum. 
Having such a non-trivial embedding, extra consistency conditions must be 
fulfilled, which we now study on a case-by-case basis. 

In the case of a $Z_2$ orbifold, $r^2=1$, $r t_i=t_i^{-1} r$, and we have
to impose these conditions also on the corresponding transformations of the 
spinors. Non-trivial solutions to these conditions exist~\cite{Lee:2005tk}, 
as can be easily demonstrated explicitly: The transformation associated with 
$r$ is $\tilde{R}=S(r)R$, where $S(r)$ is the phase rotation of the two 4d 
Weyl spinors coming from a {\bf 4} of SO(1,5). In the $Z_2$ case, we have 
$S(r)=i{\mathbbm 1}$. Choosing $R=\mbox{diag}(-i,i)$, we find $\tilde{R}= 
\mbox{diag}(1,-1)$.\footnote{The computation above can be generalized
to the case of the scalars present in a hypermultiplet coming e.g. from
a 10d gauge multiplet, which form a doublet of SU(2)$_R$ and also a doublet
of SU(2)$_{R'}$.
There is no direct action of the rotation on the scalars, which therefore
transform only due to the embedding of $r$ in SU(2)$_R\times$ SU(2)$_{R'}$
via $\tilde{R}=R\otimes R^\prime$.
Given $R$ as above, $R'$ must be chosen such that 
$\tilde{R}^2={\mathbbm 1}$, e.g. $R^\prime= R$.}
This matrix satisfies the required consistency relations 
with $T_i=\exp(i\alpha_i\sigma_2)$. In case only one of the $T_i$'s is non 
trivial, e.g. $\alpha_6=0$ and $\alpha_5=\alpha$, we can shrink the $x^6$ 
direction, obtaining a 5d effective field theory compactified on $S^1/Z_2$. 
In this case it is well known that the continuous SS parameter $\alpha$
can be described by a tunable constant superpotential $W_0\sim\alpha$~\cite{
Dudas:1997jn}. In the rest of the paper, we mainly consider such a $T^2/Z_2$ 
compactification, the 4d field content of which was already anticipated in 
Sect.~2. Notice that with such a field content a constant $W_0$ leads, in 
absence of any other effects, to SUSY breaking with zero tree-level 
potential, as expected in a SS reduction.

In case of a $Z_3$, $Z_4$ or $Z_6$ reduction, the field content would be even 
more appealing, since the $\tau$ multiplet is projected away. However, the 
consistency conditions for a SS reduction are now more stringent and cannot 
be satisfied, not even by discrete SS twists. To see this, let us first 
give a geometric description of SS breaking on a $T^2/Z_2$ orbifold. The 
compact space emerging after the orbifold projection has the topology of 
a sphere and contains 4 conical singularities, each with an opening angle 
$\pi$. The SU(2)$_{\rm R}$ twists create a non-trivial R symmetry 
holonomy for paths encircling any of the singularities. To avoid hard 
supersymmetry breaking at the singularities, the size of the corresponding 
SU(2)$_{\rm R}$ rotations has to match the opening angle of the conical 
singularity. This ensures that, in the local environment of each singularity, 
a covariantly constant spinor exists. Specifically, using the canonical map 
from SU(2) to SO(3), the R symmetry twist at each singularity, mapped to 
SO(3), has to be $\pi$ (matching the rotation in physical space). The overall 
SUSY breaking to ${\cal N}=0$ arises from a misalignment of the 4 twists at 
the 4 conical singularities. This is clearly possible since one can find 4 
SO(3) rotations around different axes which altogether give a trivial 
rotation. (The product of the 4 rotations has to be trivial since a path 
encircling all 4 singularities can be contracted without encountering another 
singularity.)

Now consider a $T^2/Z_3$ orbifold instead. The fundamental space still has
the topology of a sphere, but this time with 3 conical singularities, each 
having an opening angle of $2\pi/3$. The R symmetry twist at each 
singularity (when mapped to SO(3) in the canonical way) has to be $2\pi/3$
to avoid hard local SUSY breaking. Given again the global constraint (a loop
encircling all 3 singularities is equivalent to a trivial loop), we need
to find 3 rotations of magnitude $2\pi/3$ each which, when multiplied, give 
${\mathbbm 1}$. Elementary geometry shows that this is only possible when all
3 rotation axes coincide, in which case an ${\cal N}=1$ supersymmetry survives 
in the complete model. Thus, no SS breaking to ${\cal N}=0$ in 4d is possible. 
The above argument can be easily extended to the $Z_4$ and $Z_6$ cases. In 
both cases one again has the topology of a sphere with 3 conical 
singularities. The opening angles are $(\pi/2,\pi/2,\pi)$ and $(\pi/3, 
2\pi/3,\pi)$ respectively. Three such rotations can not give ${\mathbbm 1}$ 
in total unless their rotation axes coincide, which again leads to 
${\cal N}=1$ in 4d. 

Of course, it is also possible to obtain contributions to $W_0$ by 
introducing SS twists along some of the 4 hidden extra dimensions of an 
underlying string model, or by considering localized effects within
the ${\cal N}=1$ sectors at the orbifold singularities
(such as brane-localized gaugino condensation).

\section{Moduli stabilization}
In this section we study the stabilization of our model. Besides
the $D$-term potential induced by the flux and the superpotential
generated by the gaugino condensate we assume a constant piece 
of superpotential which has a negative sign compared to the superpotential
from the gaugino condensate. This $W_{0}$ is crucial for the stabilization of 
the modulus $s$. We incorporate perturbative corrections in a second step.
\par

To be more precise we start with the following ingredients:
\begin{eqnarray}
K &=& -\log(T+\overline{T})-\log(S+\overline{S})-\log(\tau+\overline{\tau}),\\
W &=& \mu^{3}\exp(-aS)+W_{0},
\end{eqnarray} 
assuming for simplicity that $a$, $\mu$ and $W_{0}$ are real.
The complete scalar potential is then given by
\begin{eqnarray}
V&=&\frac{1}{st(\tau+\overline{\tau})}
\left(
\mu^{6}(a^{2}s^{2}+2as)\exp(-as)+
2W_{0}\mu^{3}as\cos\left(\frac{ac}{2}\right)\exp\left(-\frac{as}{2}\right)\right)
+\frac{f^{2}}{2st^{2}}
\nonumber\\
&=&
\frac{\tilde{V}(s)}{t(\tau+\overline{\tau})}
+\frac{f^{2}}{2st^{2}}, \label{VV}
\end{eqnarray}
where the last equation has to be read as a definition of $\tilde{V}(s)$.
This potential stabilizes both $s$ and $t$ at a negative value of $V$, as is
shown in the following.\par

Consider first the `axionic' partner of $s$, denoted by $c$. As $W_{0}$ is 
taken to be negative, while $a$ and $\mu^{3}$ are positive, $c$ is always 
stabilized at a value where the cosine is unity. Thus we assume $c=0$ in the 
following. Since the shift symmetry acting on the modulus $b$  (the `axionic'
partner of $t$) is gauged, $b$ is absorbed in the massive vector boson. 
Further effects have to be taken into account to stabilize the complex 
structure modulus $\tau$, for which we assume $2\tau=1$ from now on.
As explained in Sect.~4, the problem of $\tau$ stabilization does not arise
in a  $T^2/Z_n$ ($n>2$) model, where $\tau$ is projected away.
The only caveat in these cases is that a non-zero superpotential has to be
introduced either by SS twists along some of the 4 hidden extra
dimensions of an underlying string model, or by localized
effects associated with the ${\cal N}=1$ sectors at the orbifold singularities
(such as brane-localized gaugino condensation). Alternatively, $\tau$
stabilization could result from the non-trivial $\tau$ dependence of
the Casimir energy, which, for simplicity, we do not consider in our
computation (see e.g.~\cite{Ghilencea:2005vm}).
\par

To get some intuition for the stabilization of $s$ and $t$, it is 
advantageous to first set $f=0$ and $t=1$. Then the remaining modulus 
$s$ enters the potential in exactly the same fashion as in the KKLT model. 
At the minimum of the potential, $s$ has to solve $D_{S}W=0$, so that we find
\begin{equation}
W_{0}+\mu^{3}e^{-\frac{as}{2}}(1+as)=0.
\end{equation}
This is equivalent to minimizing $\tilde{V}(s)$. For small $W_{0}$ we find the
approximate solution
\begin{equation}
as_{0}\sim 2\ln(-\mu^{3}/W_{0}).
\end{equation} 

This equation shows that $as_{0}$ can be made parametrically large by tuning 
$W_{0}$ to have small negative values. As an example consider $W_{0}=-0.01$,
$\mu^{3}=10$ and $a=1$. The result is $s_{0}\sim 20$.
\par
The approximate value at which $t$ is stabilized can be found by setting 
$s=s_{0}$. This is reasonable as the extra 
$1/s$ contribution coming from the $D$-term potential will not alter the value
of $s$ at the minimum significantly. The resulting potential for $t$ is then 
\begin{equation}
V(t)=\frac{f^{2}}{2s_{0}t^{2}}+\frac{\tilde{V}(s_{0})}{t},\label{Vt}
\end{equation}
which is minimized by 
\be
t_{0}=-\frac{f^{2}}{s_{0}\tilde{V}(s_{0})}.\label{tmin}
\ee 
Equation~(\ref{VV}) implies $\tilde{V}(s_{0})\sim-10^{-5}$.
In our example we take the flux to have its minimal nonzero value. Due to the
quantization condition in the orbifold case, $f=4\pi N$, this is $4\pi$.
With these numbers  Eq.~(\ref{tmin}) gives $t_{0}\sim~10^{6}$.
The exact potential is displayed as a contour plot in Fig.~1.
\begin{figure}
\begin{center}
\includegraphics[height=7cm]{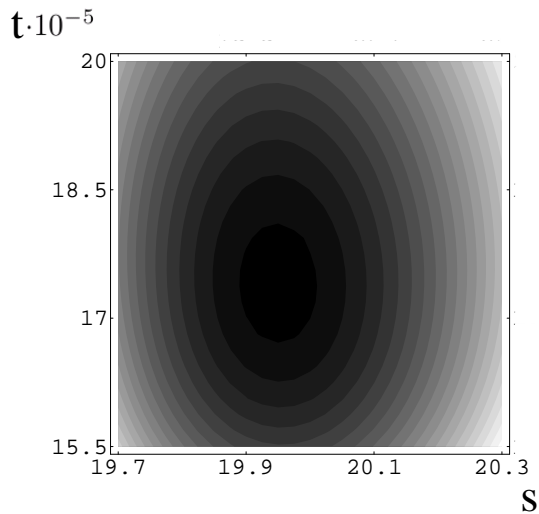}\label{contour}
\caption{$F$-term and $D$-term 
scalar potential as a function of $s$ and $t$.}
\end{center}
\end{figure}
At the minimum both $s$ and $t$ take roughly the expected values. 
It is worth noting that the minimum of the potential is always at a negative 
value in this setup, as is best seen from Eq.~(\ref{Vt}). The 
positive piece quadratic in $1/t$ is dominant for small $t$, whereas the 
negative piece linear in $1/t$ is dominant for large $t$. This tells us that 
$V(t)$ comes from positive values and approaches zero from below 
for $t\rightarrow\infty$. So clearly $V$ is negative in the minimum. This 
behavior is a result of the simple $t$ dependence of the K\"ahler potential.
\par
We now want to comment on the overall consistency of our solution. 
For the effective 4d description to be valid we need the compactification
scale to be below the 6d Plank scale. At the same time the Yang-Mills scale
has to be below the 6d Plank scale, but above the compactification scale. We
thus require the scales of our model to fulfill
$M_\text{Pl,6}^{2}>M_\text{YM,6}^{2}>M_\text{C}^{2}$.
The squared Yang-Mills scale in 6d is given by the prefactor of the 6d 
gauge-kinetic term, so it is equal to $\exp(\phi_{0})$. 
With $\exp(\phi)=\sqrt{s/t}$ we find $M_\text{YM,6}^{2}\sim 10^{-2}$, 
so that the first inequality 
holds.\footnote{Note that we have chosen units in which $M_\text{Pl,6}=1$.} 
The compactification scale is set by the volume of the internal dimensions,
$M_\text{C}^{2}=r_{0}^{-2}=(s_{0}t_{0})^{-1/2}\sim 10^{-4}$, so that the
second inequality also holds.
\par
The perturbative corrections of Sect.~3 do not alter the stabilization
qualitatively.  
As a contribution to the effective action, they can simply be added
to the scalar potential, which now reads
\begin{equation}
V(t)=\frac{f^{2}}{2st^{2}}+\frac{\tilde{V}(s)}{t}
-0.053\frac{1}{(2\pi)^{2}}\left(\frac{f}{st}\right)^{2}J_{N}^{\pm}.
\end{equation}
We see that the minimum is driven to slightly larger values of $s$ and $t$.  
It is interesting that the loop correction becomes more important than the
$D$-term potential for large fluxes.\footnote{Recall that
$J_{N}^{\pm}\simeq \vert N\vert\sim \vert f\vert/4\pi$ for large $\vert N\vert$.} 
This can be understood physically since the degeneracy of the spectrum grows with
the flux. Increasing the monopole number, and thus the flux,
is equivalent to increasing the degrees of freedom that are present on each 
Kaluza-Klein level.

\section{Conclusions}
We have approached the set of problems associated with moduli-stabilization 
and $D$-term uplift from the perspective of a simple field-theoretic model. 
Motivated by the apparent inconsistency between the flux and the gaugino
condensate, we have studied an explicit compactification of 6d supergravity, 
which allows for these two ingredients. This model is directly relevant from a 
string-theoretic perspective since it can be seen as an intermediate step in 
the compactification of 10d string theory on a ``highly anisotropic'' 
background, with 4 small and 2 large internal dimensions~\cite{orbgut}.

The gauging and the $D$-term potential that arise upon introduction of the 
flux are determined and found to match in the standard supergravity fashion, 
confirming that the flux really triggers a $D$-term. The modulus that enters 
the superpotential generated by gaugino condensation is different from the 
modulus on which the gauged shift symmetry acts. Any potential inconsistency 
is thus avoided in a natural and attractive way.

To stabilize our model, we discuss two sources for extra potential terms: an R 
symmetry twist and perturbative corrections. The R symmetry twist is 
described in terms of a constant superpotential $W_0$, so that one of the two
main compactification moduli is fixed in a fashion similar to the KKLT model. 
The other modulus is stabilized by the interplay between the $D$-term and the 
$F$-term potential. This mechanism always leads to a non-supersymmetric AdS 
vacuum in which supersymmetry is broken by both the $D$-term and the $F$-term. 

As a perturbative correction, we considered the Casimir energy of a charged 
hypermultiplet in the presence of flux. We explicitly calculate these 
loop corrections for both the $T^2$ and $T^2/Z_2$ geometry. From the 
supergravity perspective, they can be viewed as K\"ahler corrections, which 
we also display explicitly. In many cases, our corrections will be more 
important than the vacuum energy induced by the Scherk-Schwarz twist, since 
the latter becomes parametrically small in the limit of small $W_0$. By 
contrast, the flux-induced corrections can not be tuned to be small because 
of flux quantization. It would be interesting to find the counterpart of 
string-theoretic $\alpha'$ corrections in our 6d framework and to compare 
them to the flux-induced Casimir energy. 

The above perturbative corrections do not destabilize the non-SUSY AdS
vacuum we found previously. However, they are also unable to provide the 
desired uplift. Thus, a phenomenologically relevant construction would have 
to include further effects, such as an $F$ term potential arising in the 
${\cal N}=1$ sectors localized at the orbifold fixed points.

%\section*{Acknowledgements}
\acknowledgments
It is a pleasure to thank Felix Br\"ummer, Stefan Groot Nibbelink,
Boris K\"ors, Christoph L\"udeling, Dieter L\"ust, Marek Olechowski,
Claudio Scrucca, Stephan Stieberger, Riccardo Rattazzi,
Giovanni Villadoro, and Fabio Zwirner for comments and discussions.

\begin{appendix}

\section*{Appendix: Computation of the Casimir energy}
To compute the Casimir energy we use that
\be
\int \frac{d^{D}\hspace{-.2ex}k}{(2\pi)^{D}}\ln\left(k^{2}+m^{2}\right)
=-\frac{\Gamma(-D/2)}{(4\pi)^{D/2}}m^{D}.
\ee
We need to compute expressions of the form
\be
\sum_{n}\int \frac{d^{D}\hspace{-.2ex}k}{(2\pi)^{D}}\ln
\left(k^{2}+m_{n}^{2}(\delta)\right)=
-\frac{\Gamma(-D/2)}{(4\pi)^{D/2}}
\sum_{n}m_{n}^{D}(\delta)\equiv I_{\delta},
\ee
where 
\be
m_{n}^{2}(\delta)=\frac{2\vert f\vert}{r^{4}}\left(n+\tfrac{1}{2}+\delta\right)
\ee
are the bosonic/fermionic spectra as computed in Sect.~2.
Using the Hurwitz and Riemann zeta functions, denoted by $\zeta_{H}$ and 
$\zeta_{R}$ respectively \cite{Elizalde:1994gf}, we find that
\be
I_{\delta}=-\frac{\Gamma(-D/2)}{(4\pi)^{D/2}}\left(\frac{2\vert f\vert}
{r^{4}}\right)^{D/2}\zeta_{H}(-D/2,\delta+\tfrac{1}{2}).
\ee 
The limit $D\rightarrow 4$ for $\delta=\pm 1/2$ and $\delta=0$, which
are the cases of interest to us, can be computed by noting that 
$\zeta_{R}(-2)=0$ and using the expansions 
\bea
\Gamma(\epsilon-2)&=&\frac{1}{2\epsilon}+\mathcal{O}(1)\\
\zeta_{H}(\epsilon-2,1/2)&=&(2^{\epsilon-2}-1)\zeta_{R}(\epsilon-2)
=-\frac{3}{4}\zeta_{R}'(-2)\epsilon+\mathcal{O}(\epsilon^{2}).
\eea
We find that
\bea
I_{0}&=&\frac{3}{8}\frac{1}{(4\pi)^{2}}\left(\frac{2f}{r^{4}}\right)^{2}
\zeta_{R}'(-2)\\
I_{1/2}=I_{-1/2}&=&-\frac{1}{2}\frac{1}{(4\pi)^{2}}
\left(\frac{2f}{r^{4}}\right)^{2}
\zeta_{R}'(-2),
\eea
where $\zeta_{R}'(-2)=-\zeta_{R}(3)/(4\pi^{2})=-0.0304$. The 
equality in the last line follows from 
$\zeta_{H}(x,1)=\zeta_{H}(x,0)\equiv\zeta_{R}(x)$.
\subsection*{The $T^{2}$ case}
Taking the degeneracy of the spectra and the flux quantization ($f=2\pi N$) 
into account, the Casimir energy of one charged hypermultiplet on $T^{2}$ can 
be expressed as
\bea
V_{\mbox{\scriptsize{Casimir}}}&=&2\vert N\vert I_{0}-2\vert N\vert I_{1/2}\nonumber\\
&=&\frac{7}{4}\frac{\vert N\vert^{3}}{(st)^{2}}\zeta_{R}'(-2)\cong-0.053
\frac{1}{(2\pi)^{3}}\frac{\vert f\vert^{3}}{(st)^{2}}
\eea

\subsection*{The $T^{2}/Z_{2}$ case}
To find which states are projected away in the orbifold case we need to 
determine the parity of the zero eigenfunctions.
Up to normalization they read
\be
\Phi_{j}=\sum_{m=-\infty}^{\infty}
\exp\left(-\tfrac{1}{2}\vert f\vert\left(x_{5}-\frac{1}{\vert 2 N\vert}
(\vert 2 N\vert m+j)\right)^{2}\right)
\exp\left(2\pi i(\vert 2 N\vert m+j)x_{6}\right),
\ee
where we have used an appropriate gauge\cite{Bachas:1995ik} and imposed the
$T^2/Z_2$ flux quantization condition $f=4\pi N$.
By shifting $m$ 
it is easy to see that $\Phi_{j}=\Phi_{j+\vert 2 N\vert}$, so that there are 
$\vert 2 N\vert$ distinct eigenfunctions.  We find that the parity operation 
(i.e. the $Z_{2}$ rotation) maps $\Phi_{j}$ to $\Phi_{-j}$ and hence to 
$\Phi_{\vert 2 N\vert-j}$. 
Thus we conclude that 
\be
\Phi_{j}^{e}\equiv\Phi_{j}+\Phi_{\vert 2 N\vert-j}
\ee
has even parity and
\be
\Phi_{j}^{o}\equiv\Phi_{j}-\Phi_{\vert 2 N\vert-j}
\ee
has odd parity. Note that there is no $\Phi_{\vert N\vert}^{o}$,
but just a $\Phi_{\vert N\vert}^{e}=2\Phi_{\vert N\vert}$. Furthermore 
$\Phi_{\vert 2 N\vert}$ always has even parity. 
Besides these exceptions the rest of the spectrum pairs up according to the 
equations above. The number of even eigenfunctions ($N_{e}$) is 
then
\bea 
N_{e}&=&\vert N\vert+1.
\eea
To find the number of remaining states on the excited levels, we use
the fact that the raising and lowering operators are linear in 
$\mathcal{D}_{5}$ and $\mathcal{D}_{6}$, so that they anticommute with the 
generator of the $Z_{2}$. 
\par
The two complex bosons have different internal parity assignments, so that 
we find $\vert 2 N\vert$ of them on each mass level.
This is not true for the fermions, 
because the ground states of different chirality, and hence different internal
parity, have different masses:
\be
(m_{n}^{2})_{\pm}=\frac{2\vert f\vert}{r^{4}}
\left(n+\tfrac{1}{2}\pm\tfrac{1}{2}\right).
\ee
We first analyse the tower containing massless states and assume that its 
fermions have positive internal parity.\footnote{If the massless fermions have
negative internal parity, the computation is the same with $N_{e}$ and 
$(\vert 2N\vert-N_{e})$ interchanged.}
On the ground state we find $N_{e}$ massless fermions that remain in the
spectrum. By acting $2n$ times with the raising operator we find $N_{e}$
surviving states on the level $2n$, so that we generate a spectrum with
masses
\be
m_{n}^{2}=\frac{2\vert f\vert}{r^{4}}(2n)
\ee
and degeneracy $N_{e}$.
If we consider the $\vert 2N\vert-N_{e}=\vert N\vert -1$ states that are
projected away from the ground state and act once with the raising operator,
we find $\vert N\vert -1$ fermions on the first excited level that are even
under the orbifold projection.\footnote{Note that the raising operators do
not change the chirality.} From there we can again act $2n$ times with the
raising operator to find more states that remain in the spectrum. We thus
generate a second tower with masses
\be
m_{n}^{2}=\frac{2\vert f\vert}{r^{4}}(2n+1)
\ee
and degeneracy $\vert N\vert-1$.
We now turn to the fermions of the opposite chirality and hence opposite
internal parity. On the ground state of this tower we find
$\vert 2N\vert-N_{e}=\vert N\vert -1$ remaining states with masses
$2\vert f\vert/r^{4}$. By the same argument as 
above this yields a spectrum
\be
m_{n}^{2}= \frac{2\vert f\vert}{r^{4}}(2n+1),
\ee
with degeneracy $\vert N\vert-1$.
Acting with the raising operator once on the $N_{e}$ ground states that are
projected away we find $N_{e}$ states on the first excited level that remain
in the spectrum. These generate a tower of masses
\be
m_{n}^{2}=\frac{2\vert f\vert}{r^{4}}(2n+2)
\ee
with degeneracy $N_{e}$. 
As expected, the degeneracy of each state is roughly half of what
we found before performing the $Z_{2}$ projection. 
\par
By appealing to the definitions made at the beginning of the Appendix we find 
\be
V_{\mbox{\scriptsize{bosons}}}=\vert 2N\vert I_{0},
\ee
and
\be
V_{\mbox{\scriptsize{fermions}}}^{+}
=-8\left((\vert N\vert+1)I_{-1/2}+(\vert N\vert-1)I_{0}\right)
\ee
if the massless fermions have positive internal parity. If the
massless fermions have negative internal parity the fermionic
contribution to the Casimir energy reads:
\be
V_{\mbox{\scriptsize{fermions}}}^{-}
=-8\left((\vert N\vert-1)I_{-1/2}+(\vert N\vert +1)I_{0}\right).
\ee
Putting everything together and using the explicit expression
for $N_{e}$ the complete Casimir energy reads
\be
V_{\mbox{\scriptsize{Casimir}}}^{\pm}=7\left(\frac{N}{st}
\right)^{2}\zeta_{R}'(-2)J_{N}^{\pm},
\ee
where we have defined
\be
J_{N}^{\pm}\equiv \vert N\vert\pm 4.
\ee
The different signs in $J_{N}^{\pm}$ are related to the different parities 
of the massless fermions: if the parity is positive, the sign is `$+$', if the 
parity is negative, the sign is `$-$'. Note that we recover the Casimir energy 
of the untruncated spectrum if we add $V^{+}$ and $V^{-}$ and remember that
$N=f/4\pi$.

\end{appendix}

\end{document}